# Evolving E-commerce Logistics Planning: Integrating Embedded Technology and Ant Colony Algorithm for Enhanced Efficiency

**Lynn Huang[1*]**

[1]*Sawyer business school, Suffolk university, Boston, MA, USA*

[*]*Corresponding Author*

**Abstract:** Amidst the era of networking, the e-commerce sector has undergone notable expansion, notably with the advent of Cross-border E-commerce (CBEC) in recent times. This growth trend persists, necessitating robust logistical frameworks to sustainably support operations. However, the current e-commerce logistics paradigm faces challenges in meeting evolving user demands, prompting a quest for innovative solutions. This research endeavors to address these complexities by undertaking a comprehensive analysis of CBEC logistics models and integrating embedded technology into logistical frameworks, resulting in the development of an advanced logistics tracking system. Moreover, employing the ant colony algorithm, the study conducts experimental investigations into optimizing logistics package distribution route planning. Noteworthy enhancements are observed in key metrics such as average delivery time, signaling the efficacy of this approach. In essence, this research offers a promising pathway towards optimizing logistics package distribution routes and bolstering package transportation efficiency within the CBEC domain.

**Keywords:** E-commerce, Logistics, Package Tracking System, Optimization, Route Planning, Meta-heuristics

## 1. Introduction

With the advent of the network era, numerous emerging industries are witnessing rapid growth. Among these burgeoning sectors, e-commerce stands out as a prominent player experiencing substantial development [20]. Concurrently, to extend the scope of e-commerce, Cross-border E-commerce (CBEC) has emerged, exhibiting its own expanding scale of development. As e-commerce flourishes, the logistics sector has also experienced a surge in prominence, giving rise to CBEC logistics. However, the evolution of CBEC logistics has encountered various new challenges that significantly impede its progress. Consequently, addressing these challenges necessitates a multifaceted approach to actively seek solutions.

In recent years, scholarly attention has increasingly focused on the domain of e-commerce logistics, with researchers launching various studies in this field [1]. Wu and Lin (2018) [2], for instance, explored e-commerce logistics business models using unstructured big data and devised a mixed content analysis model. Empirical findings from this model, integrating resource dependence theory and innovation diffusion theory, facilitated the generation of logistics strategies. Similarly, Giuffrida et al. (2021) [3] conducted a survey of online exporters and third-party freight e-commerce logistics service providers. Through structural equation model analysis, the study examined the potential relationship between risk management strategies in e-commerce logistics and types of uncertainty. Furthermore, Wang et al. (2021) [4] proposed a novel method to delineate customer utility by incorporating customer time preference and devised three CBEC logistics models. Theoretical and numerical analyses indicated that the Online-To-Offline (O2O) mode was preferable for retailers over offline modes, with

Door-To-Door (D2D) mode proving advantageous under specific conditions. Additionally, Zhang et al. (2021) [5] tackled vehicle routing challenges in the Business-to-Consumer (B2C) e-commerce logistics system for simultaneous pickup and delivery from multiple warehouses within a specified time range. The study also developed a mixed integer nonlinear programming model and tested it across various scenarios. Nonetheless, while these scholars contributed valuable insights to e-commerce logistics, further comprehensive research remains imperative.

Divergent perspectives and studies on e-commerce logistics have also emerged among other scholars. Zheng et al. (2020) [6], for instance, utilized the Analytic Hierarchy Process and entropy method to explore logistics distribution selection modes for e-commerce enterprises, offering effective management suggestions based on experimental outcomes. Tsang et al. (2021) [7] proposed a multi-temperature e-commerce logistics distribution planning system leveraging the Internet of Things, embedding a two-stage multi-objective genetic algorithm optimizer to enhance order processing capabilities and maintain customer satisfaction levels. Addressing logistics challenges, Sachan et al. (2020) [8] introduced an anti-predatory method to predict optimal warehouse establishment for logistics network optimization. Experimental evaluations demonstrated superior results compared to genetic algorithms. Additionally, Liu et al. (2021) [9] outlined the fundamental framework of factors influencing the organizational efficiency of the intelligent e-commerce logistics ecological chain, emphasizing technological innovation capability and contract performance as crucial determinants. Overall, while research in e-commerce logistics remains relatively scarce, there is a pressing need to enhance research on e-commerce logistics package tracking and integrate embedded technology into logistics systems.

This paper conducts a comprehensive analysis of the CBEC logistics model and proposes solutions to its deficiencies. By establishing a comprehensive intelligent logistics information system, implementing various overseas warehouse management strategies, and adopting a personalized supply chain service mode, the shortcomings of CBEC logistics can be mitigated. Additionally, the paper develops a logistics tracking system based on embedded technology to ensure package transportation security and enhance logistics services, thereby augmenting users' shopping experiences.

## 2. Embedded Technology, CBEC Logistics and Package Tracking System

*2.1 Embedded Technology*

Embedded technology, a pivotal component of computer systems and equipment, finds widespread application across various domains. Typically, any device featuring digital interfaces utilizes embedded technology, with some embedded systems also incorporating operating systems. Encompassing electronic devices, hardware configuration, software, Internet connectivity, multimedia capabilities, and more, embedded technology boasts versatile applications in numerous fields.

In the realm of e-commerce, embedded technology plays a significant role. Within online shopping platforms, embedded technology, along with data mining and wireless communication technologies, facilitates the collection of shopping information from a diverse user base, spanning different geographical regions. Given the varied shopping behaviors and unpredictable shopping times of users, the integration of data mining technology becomes crucial for deep analysis and extraction of valuable user data. This enables the identification of potential target users and enhances the overall shopping experience by providing personalized intelligent shopping guidance services.

*2.2 CBEC Logistics*

*2.2.1 Definition of CBEC Logistics*

Cross-border E-commerce (CBEC) logistics refers to the logistical operations arising from enterprises and consumers engaging in cross-border e-commerce transactions across different nations [10]. The seamless flow of products is facilitated by connecting logistics systems within and outside the country. CBEC logistics encompasses domestic freight, international logistics, and destination country freight, heavily relying on e-commerce platforms for storage, delivery, transportation, and circulation support.

Currently, a significant portion of CBEC transactions is export-oriented. Business-To-Business (B2B) model exports have minimal impact on the logistics industry, resembling traditional physical transactions with large quantities, stable operations, and low per-product freight costs. However, CBEC logistics associated with export-oriented Business-To-Consumer (B2C) and Consumer-To-Consumer (C2C) models are more complex.

CBEC logistics involves navigating various uncertainties, including high costs and complex transportation processes. Traditional logistics methods, such as international postal parcels and international express services, are crucial for CBEC logistics. Additionally, cross-border special line logistics are tailored for specific countries and regions, ensuring tailored delivery but requiring strict adherence to transit times.

Emerging logistics strategies include overseas warehousing, where exporters pre-transport goods to overseas leased warehouses, enhancing the shopping experience by reducing freight costs and transit times. Third-party logistics services facilitate goods transportation without participating in product trading, ensuring prompt product delivery and comprehensive service offerings at lower costs for higher product volumes.

*2.2.2 Limitations in the Advancement of CBEC Logistics*

The nature of CBEC transactions involves cross-border product exchanges, introducing complexities inherent to international trade affairs. The transaction process is inherently intricate, comprising numerous transactional links. Logistics transportation spans multiple countries, each with its own distinct legal and regulatory frameworks, engendering numerous uncontrollable factors in the transaction process. Consequently, logistics incur various indirect costs during transportation, necessitating additional expenditures such as tariffs, loss fees, and other associated risks. Significant investments in manpower, material resources, and capital are indispensable for product transportation, culminating in a pronounced escalation of logistics costs, thereby exerting adverse effects on CBEC logistics.

The protracted transportation duration, driven by cost minimization imperatives, prompts many businesses to opt for postal parcel transportation. However, the efficiency of this mode is suboptimal, frequently resulting in excessively prolonged product transit times and sluggish delivery speeds, thereby compromising the overall shopping experience for customers. Moreover, the protracted transportation duration poses challenges for buyers seeking to return or exchange goods. Lengthy delivery times exacerbate the waiting period for buyers to receive the goods, deterring many consumers from pursuing returns or exchanges due to the associated time commitment.

*2.2.3 Strategies for Advancing the CBEC Logistics Framework*

In order to foster the advancement of CBEC logistics, several strategies can be implemented. Creating a complete intelligent logistics information system: CBEC logistics must evolve towards intelligence, digitalization, and informatization. This entails establishing a smart logistics information system that leverages emerging technologies to enhance transportation, delivery, warehousing, and other services [11][12]. Strengthening internet infrastructure is vital for the stable operation of transport management systems. Continuous improvement of the cross-border logistics management system is essential to optimize business scale, transportation efficiency, and overall logistics services. Scientific planning of cross-border logistics park structures, along with optimization of existing facilities and enhanced park management levels, further contributes to improved daily operations through the application of artificial intelligence technology.

Managing overseas warehouses in various ways: Overseas warehouses play a crucial role as platforms for CBEC logistics services. They must cater to diverse user purchase behaviors and regional influences by strengthening localized services. Expanding logistics service offerings and improving localization can provide users with a broader range of services. Leveraging big data technology enables comprehensive exploration of purchase preferences and consumption behaviors among different user groups, thereby enhancing efficient product allocation management in overseas warehouses. Integration of various logistics modes, such as border warehouses, overseas warehouses, and logistics special lines, enables a comprehensive and adaptable logistics approach to meet the rapidly changing cross-border market demands.

Developing a customized supply chain service mode: CBEC full-chain management focuses on specialized projects, with logistics enterprises offering a range of supply chain management services to e-commerce entities. The goal is to provide complete full-chain services while controlling logistics costs [13]. Developing customized supply chain service models effectively enhances the import and export competitiveness of high-value products with long industrial chains. Products operated through CBEC have varying grade-specific cross-border logistics requirements, with tailored international supply chain services catering to special products, high-end products benefiting from international air logistics, and medium to low-grade daily products utilizing international general logistics for cost-effective transportation.

*2.3 Embedded Technology-Integrated E-commerce Logistics Tracking System*

The logistics tracking system comprises two primary components: the information subsystem and the operation subsystem. The former facilitates various services such as loading, unloading, delivery, commodity circulation, storage, and packaging, while the latter supports functions related to delivery, ordering, and dispatch [14]. Operationally, the logistics tracking system coordinates these subsystems to ensure efficient logistics transportation, cost control, and service level enhancement. Management-wise, its objective is to provide users with efficient transportation and after-sales services.

Embedded systems represent a novel technology grounded in electronic information technology, resembling other microprocessors in its hardware and software components. The hardware section encompasses an embedded microprocessor, memory chip, and output device, while the software integrates command and data functions. Embedding technology into the logistics tracking system offers numerous advantages, notably in reducing freight consumption during transportation, thereby significantly enhancing logistics speed and providing users with expedited logistics services.

The module design of the e-commerce logistics tracking system centers on key technologies such as the Internet of Things (IoT), supplemented by sensors, wireless communication, and embedded technology to monitor package statuses effectively. The system comprises three primary modules: sensor module, embedded onboard module, and service terminal module. Sensor module incorporates a single-chip computer, sensor, and wireless transmission module to relay package status information to the onboard embedded equipment via wireless transmission. Embedded onboard module: Comprising an embedded chip, Global Positioning System (GPS), General Packet Radio Service (GPRS) module, and barcode scanner, this module facilitates the storage of package information via the barcode scanner, GPS acquisition of package location information, and wireless reception of various package data. The service terminal module comprises essential components such as a server, database, and map. Specifically, data from the embedded vehicle terminal is stored within the database. The Active Server Pages (ASP) web page accesses the database information, allowing for real-time querying of package transportation details and location information via the map interface.

When a package experiences sudden events such as collision, disassembly, or damage, the sensor equipment utilizes the wireless module to transmit package information to the embedded vehicle module. Subsequently, the vehicle system forwards this information to the network server for storage, ensuring package safety. The operational procedure of the package sensor device is detailed in the accompanying illustration. Within the vehicle terminal module, logistics packages are typically scanned and identified using barcodes. Additionally, GPS assists in acquiring package location information, which is then transmitted to the vehicle terminal module. Here, it is automatically integrated into the database system alongside data collected by the sensor equipment. The operational workflow of the service terminal module is elucidated in the provided visual aid.

## 3. E-business Logistics Terminal Distribution Route Planning Based on Ant Colony Algorithm

*3.1 Mathematical Formulation for Determining the Location of the Logistics Distribution Center*

The location of the logistics distribution center needs to comprehensively consider the logistics resources and market demand [15]. At the same time, it should adhere to the principles of economic applicability, convenient transportation, coordination and optimization, and take comprehensive consideration to select the most appropriate delivery address.

The vehicle drives out of the distribution center and passes through multiple target customer points before return the original distribution center in the end. The mathematical model could be expressed as follows:

Sets:
- N = set of customers (n = 1, 2, ..., |N|)
- M = set of distribution center (m = |N|+1, ..., |N|+|M|)
- V = set of vehicles (v = 1, 2, ..., |V|)
- dij = distance or travel cost between point i and j

Decision Variables:
- $x_{vij}$ = 1 if vehicle v travels from point i to point j, 0 otherwise
- $y_{vi}$ = 1 if vehicle v leaves distribution center i, 0 otherwise

Objective function:

$$\min w_1 \sum_{i=|N|+1}^{|N|+|M|} 1_{\sum_{v=1}^{|V|} y_{vi} \geq 1} + w_2 \sum_{v=1}^{|V|} \sum_{i=1}^{|M|+|N|} \sum_{j=1}^{|M|+|N|} d_{ij} * x_{vij} \quad (1)$$

Subject to:

$$\sum_{v=1}^{|V|} \sum_{i=1}^{|M|+|N|} x_{vij} = 1, j = 1,2,\cdots,|N| \quad (2)$$

$$\sum_{v=1}^{|V|} \sum_{j=1}^{|M|+|N|} x_{vij} = 1, i = 1,2,\cdots,|N| \quad (3)$$

$$\sum_{j=1}^{|N|} x_{vmj} = \sum_{i=1}^{|N|} x_{vim} \leq 1, v = 1,2,\ldots,|V|, \ m = |N|+1,\cdots,|N|+|M| \quad (4)$$

$$\sum_{i=|N|+1}^{|M|+|N|} y_{vi} = 1_{\sum_{i,j}^{|M|+|N|} x_{vij} \geq 1}, v = 1,2,\cdots,|V| \quad (5)$$

$$\sum_{j=1}^{|N|} x_{vij} = y_{vi}, v = 1,2,\cdots,|V| \quad (6)$$

$$x_{vij}, y_{ki} \subseteq \{0,1\} \quad (7)$$

The objective function encompasses a weighted average of two components: the first component pertains to the aggregate count of distribution centers, while the second component strives to minimize total distances. Constraints (2) and (3) delineate flow balance restrictions for each customer point, mandating that each point be visited precisely once. Constraint (4) imposes flow balance constraints on the distribution center candidates. Constraint (5) stipulates that a vehicle must commence its route from a distribution center if it is utilized, while constraint (6) facilitates the updating of variables to preclude the formation of cycles among customer points. Constraint (7) delineates binary constraints applicable to all variables.

*3.2 Ant Colony Algorithm Calculates the Shortest Path*

Metaheuristics are algorithmic strategies for solving complex optimization problems where exact methods are impractical. They iteratively explore solution spaces, leveraging randomness and heuristic principles to efficiently approximate optimal solutions [16]. Examples include genetic algorithms, honey badger algorithm [17], and ant colony optimization [21], widely applied across diverse disciplines for tackling challenging optimization tasks [19]. The Ant Colony (AC) algorithm constitutes a metaheuristic optimization technique inspired by the foraging behavior observed in real ant colonies. In their natural environment, ants adeptly navigate the shortest path from their nest to a food source by depositing pheromones, effectively marking their route. The algorithm iteratively generates solutions by simulating this process, where artificial ants traverse from one node to another based on a probabilistic rule. Notably, the probability of selecting a specific edge is influenced by the concentration of pheromone along that edge. Evaluation of solution quality occurs at the conclusion of each iteration. Leveraging the Ant Colony algorithm facilitates the identification of near-optimal shortest distances for distribution routes, thereby facilitating route planning. Within this framework, artificial ants serve as vehicles, tasked with devising distribution paths for each designated target point.

It is assumed that the transfer probability of the ant with number $o$ from one target point to another is as follows:

$$u_{k,i}^o = \begin{cases} a_i \frac{\varphi_{k,i}^\gamma \cdot \omega_{k,i}^\varepsilon \cdot \sigma_{k,i}^\theta \cdot \mu_{k,i}^\rho}{\sum_{o \in S}(\varphi_{k,i}^\gamma \cdot \omega_{k,i}^\varepsilon \cdot \sigma_{k,i}^\theta \cdot \mu_{k,i}^\rho)} \\ 0, otherwise \end{cases} \quad (8)$$

Among them, $\varphi_{k,i}^\gamma, \sigma_{k,i}^\theta, \mu_{k,i}^\rho$ are prior proxy information representing for distance, freight cost, and distribution cost, respectively. $\varepsilon, \gamma, \theta, \rho$ are adjustable parameters to leverage among different factors. Pheromone trail $\omega_{k,i}^\varepsilon$ can be updated at each iteration by Formula (9):

$$\omega_{k,i} = (1-\mu) * \omega_{k,i} + \sum_{o=1}^{S} \Delta\omega_{k,i}^o \quad (9)$$

The freight proxy cost $\sigma_{k,i}^\theta$ of the distribution route at the end of e-commerce logistics includes fixed cost and transportation cost [18]. The total cost can be expressed as follows:

$$\sigma = \sigma_1 + \sigma_2 \quad (10)$$

Among them, $\sigma_1$ represents the fixed cost of the planned route, $\sigma_2$ represents vehicle transportation cost, The formula is as follows:

$$\sigma_1 = \sum_{x=1}^{n} h_x l_x \quad (11)$$

Among them, $l_x$ represents a fixed variable, that is, the use status of the xth vehicle, and $h_x$ represents the fixed cost of a single vehicle.

$\sigma_2$ represents the vehicle transportation cost, and the formula is expressed as follows:

$$\sigma_2 = \delta \sum_{j=1}^{m} \sum_{k=0}^{m} \sum_{x=0}^{n} v_{ki} y_{kix} \quad (12)$$

Among them, $\delta$ represents the logistics and transportation costs, and $y_{kix}$ represents the distribution task status of car $x$ from demand point $i$ to demand point $k$.

*3.4 Experimental results*

To assess the efficacy of the ant colony algorithm in planning distribution routes for e-commerce logistics terminals, this experiment endeavors to assess the efficacy of two distinct algorithms, the greedy approach and the proposed AC algorithm, within the realm of e-commerce logistics. The primary objective entails a comparative analysis of their performance in terms of delivery time across diverse scenarios, varying the number of vehicles employed. The Greedy Approach represents a conventional heuristic method, while the AC Algorithm leverages sophisticated optimization techniques inspired by the collective behavior of ant colonies. Through rigorous experimentation and meticulous data analysis, this study aims to elucidate the relative merits and shortcomings of each algorithm, thereby discerning their practical applicability and effectiveness in real-world logistics operations. By elucidating the algorithm that demonstrates superior performance in optimizing delivery routes and minimizing delivery time, this research endeavors to furnish valuable insights for the advancement of e-commerce logistics systems.

The experimental results showcase the performance disparity between the greedy approach and the AC algorithm across varying numbers of vehicles, ranging from 1 to 20. Initially, with one vehicle, the greedy approach yields a delivery time of 77.5 hours, whereas the AC algorithm achieves a notably lower time of 71.5 hours. As the number of vehicles increases, the delivery time reduces for both algorithms. However, the AC algorithm consistently outperforms the greedy approach across all instances. For instance, with 20 vehicles, the delivery time with the Greedy Approach is 15.84 hours, while the AC Algorithm remarkably reduces it to 7.02 hours. This trend underscores the superior efficiency and optimization capabilities of the AC algorithm in route planning and package delivery. The results highlight the potential of the AC algorithm to enhance logistics operations and expedite delivery processes, offering valuable insights for improving e-commerce logistics systems.

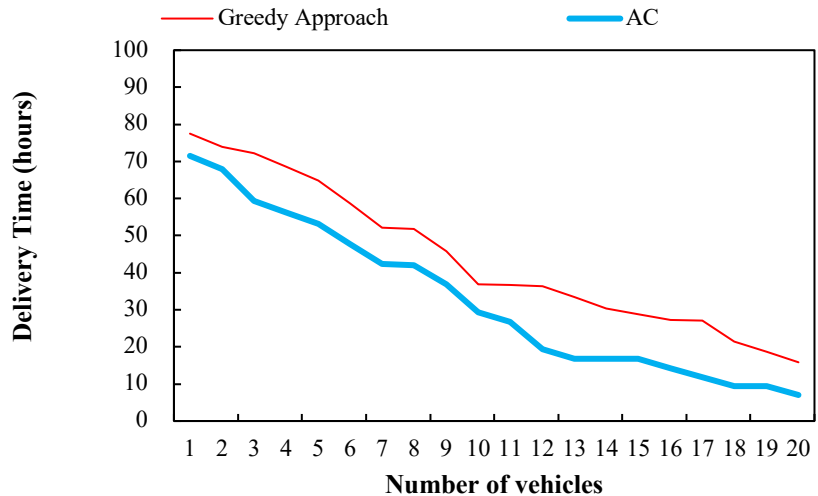

**Figure 1.** Evaluation of delivery time for logistics packages employing varied algorithms.

## 4. Conclusions

The rapid proliferation of e-commerce has precipitated the emergence of Cross-border E-commerce (CBEC), accentuating the crucial interdependence between e-commerce and logistics. With CBEC transactions steadily increasing, existing logistical infrastructures struggle to accommodate the evolving needs of consumers, underscoring the necessity for continuous innovation. To address this challenge, this study conducts an exhaustive examination of the CBEC logistics model while incorporating embedded technology to establish a comprehensive logistics package tracking system. Furthermore, the study employs the ant colony algorithm to evaluate and refine end distribution route planning within the realm of e-commerce logistics. Experimental results demonstrate the algorithm's notable reductions in logistics package delivery time, thereby optimizing delivery routes. Future research endeavors will concentrate on further refining the ant colony algorithm to better suit the practical demands of e-commerce logistics package distribution, thus enhancing its effectiveness and performance in route optimization.